\documentstyle[12pt,epsf]{article}
\def\del        {  \partial  }
\def\half       {  {1\over 2}  }

\def\defint#1#2 {  \int_{#1}^{#2}  }
\def\rootof#1   {  \left( #1 \right)^{1/2}  } 
\def\deldel#1   {  {\partial\over \partial #1}  }

\def\abs#1      {  \vert #1 \vert  }

\def\evalat#1   {  \left\vert_{#1} \right. } 
\def\e          { {\rm e}  }

\def\lsim    {\lower .65ex \hbox{\ $\stackrel{<}{\sim}$\ } }
\def\gsim    {\lower .65ex \hbox{\ $\stackrel{>}{\sim}$\ } }

\def\calO       { {\cal O} }

\def\vecii#1#2      {  \left(\begin{array}{c}#1\\#2\end{array}\right)  }
\def\veciii#1#2#3   {  \left(\begin{array}{c}#1\\#2\\#3\end{array}\right)  }
\def\veciv#1#2#3#4  {  \left(\begin{array}{c}#1\\#2\\#3\\#4
                                 \end{array}\right)  }
\def\vecv#1#2#3#4#5 {  \left(\begin{array}{c}#1\\#2\\#3\\#4\\#5
                                 \end{array}\right)  }

\def\matrixii#1#2#3#4            {  \left(\begin{array}{cc}#1&#2\\#3&#4
                                       \end{array}\right) }
\def\matrixiii#1#2#3#4#5#6#7#8#9 {  \left(\begin{array}{ccc}#1&#2&#3\\
                                     #4&#5&#6\\#7&#8&#9\end{array}\right)  }
\def\mativ#1#2#3#4               {  \left(\begin{array}{cccc}
                                       #1\\#2\\#3\\#4\end{array}\right) }
\def\matv#1#2#3#4#5              {  \left(\begin{array}{ccccc}
                                     #1\\#2\\#3\\#4\\#5\end{array}\right)  }
\def\eqabegin         {  \begin{eqnarray}  }
\def\eqaend           {  \end{eqnarray}  }
\def\nn               {  \nonumber  }
\def\bracetwo#1#2     {  \left\{ \begin{array}{l} #1 \\ #2 \end{array}
                         \right.  }
\def\bracetwocases#1#2#3#4  {   \left\{ \begin{array}{ll} #1 &
                                 \qquad #2 \\
                                 #3 & \qquad #4 \end{array} \right.  }
\def\bracebegin#1     {  \left\{ \begin{array}{#1}   }
\def\braceend         {  \end{array}\right.   }
\def\parn              {  \par\noindent }
\def\parbigskip        {  \par\bigskip  }
\def\parmedskip        {  \par\medskip  }
\def\parsmallskip      {  \par\smallskip  }

\def\parag#1           {\paragraph{#1} \mbox{ }\parmedskip\noindent}
\def\boxit#1#2      {  \vbox{\hrule\hbox{ \hskip -4.1pt \vrule\kern3pt 
                       \vbox
                    {  \hsize #1 \strut\kern3pt #2 \kern3pt\strut  }
                       \kern3pt  \vrule} \hrule  } }
\def\centerbox#1#2  {  \mbox{  }\par\bigskip  \hfil \boxit{#1}{#2} \hfil
                       \par\bigskip\noindent }
\def\rightbox#1#2   {  \hfill\boxit{#1}{#2}  }
\def\leftbox#1#2    {  \boxit{#1}{#2}  }
\def\fullbox#1      {  \boxit{\textwidth}{#1}  }
\def\trianglemap#1#2#3#4#5#6  {   {\large $$ \begin{array}{rcl} #1\!\!\!
                                  &{\stackrel{{\scriptstyle #2}}{
                              \longrightarrow   }}&\!\!\!  #3 \\ 
                            { } & {\scriptstyle #4}\!\!\!\searrow \quad
                                \swarrow \!\!\!{\scriptstyle #5}& { } \\
                                  { } & #6 & { } \end{array} $$ }    }
\def\squaremap#1#2#3#4#5#6#7#8    { {\large $$ \begin{array}{ccc}#1 &
                   \stackrel{{\scriptstyle #2}}{\longrightarrow} & #3 \\
                     {\scriptstyle #4}\!\downarrow & { } & \downarrow \!
                     {\scriptstyle #5}\\ #6 &\!\!
                      \longrightarrow_{{ }_{\!\!\!\!\!\!\!\!\!\!\!
                      {\scriptstyle #7}}}    &#8 \end{array} $$ }   }
\def\righttrianglemap#1#2#3#4#5#6  {  {\large $$ \begin{array}{rcl}
                 #1\!\! & \stackrel{{\scriptstyle #2}}{\longrightarrow} 
                      & #3 \\  { }&\!\!{\scriptstyle #4}\!\!\searrow
                      & \downarrow \!\!{\scriptstyle #5}\\
                      { }&{ }& #6 \end{array} $$ }   }
\def\rightfigspacebegin  {  \par\noindent\begin{minipage}[t]{10cm}  }
\def\rightfigspaceend    {  \end{minipage}\par\noindent  }
\def\leftfigspacebegin   {  \par\noindent
                             \hspace*{10cm}\begin{minipage}[t]{6cm} }
\def\leftfigspaceend     {  \end{minipage}\par\noindent  }
\def\titleandfile#1#2   {  \begin{center}{\Large\bf #1}\end{center}
                            \par\begin{flushright} #2 \end{flushright}  }
\def\msection#1      {  \begin{center} \section{#1} \end{center}   }
\def\nsection#1      {  \let\boldface\bf \def\bf{} \section{#1}
                           \let\bf\boldface   }
\def\mnsection#1     {  \begin{center} \nsection{#1} \end{center}  }
\def\capsection#1    {  \let\boldface\bf \def\bf{\sc} \section{#1}
                           \let\bf\boldface   }
\def\mcapsection#1   {  \begin{center} \capsection{#1} \end{center} }
\newcommand{\nullify}[1]{}
\def\period{\, .}
\def\comma{\, ,}
\def\ff{\gamma^2}  % front factor
\def\xplus{{x^+}}
\def\xminus{{x^-}}
\def\yplus{{y^+}}

\def\delplus{\del_+}
\def\delminus{\del_-}

\def\ket#1{\mid #1 >}
\def\bra#1{< #1 \mid }

\def\psmearvac{\mid \tilde{P} >} %sato
\def\psmearnorm{ < \tilde{P} \mid  \tilde{P} >} %sato
\def\Psizeroket{\mid \Psi_0 >}%sato
\def\weight{W(p^+)}%sato
\def\Ltot{L^{tot}}

\def\pplus{p^+} \def\pminus{p^-}
\def\qplus{q^+} \def\qminus{q^-} %added 8/16

\def\etap{\eta^+}

\def\At{\tilde{A}}

\def\Atil{\tilde{A}}

\def\Ltot{L^{tot}}
\def\Lbartot{\bar{L}^{tot}}
\def\dbar{{\bar{d}}}
\def\bbar{{\bar{b}}}

\def\calLm{{\cal L}_-}
\def\calLp{{\cal L}_+}

\def\gt{\tilde{\gamma}}

\def\chia{\chi_a}

\def\nchiepsi{:\chi\e^{-\psi}: }
\def\At{\tilde{A}}
\def\etap{\eta^+}

\def\Omegaket{\mid \Omega >}
\def\Omegabra{< \Omega \mid }
 
\def\Fii{{}_1F_1}

\def\xilarge{&\stackrel{\xi\rightarrow \infty}{\sim}& }
\def\cx#1#2{ (1-x^2)^{#1} (1+x^2)^{#2} }

\def\lm{\lambda_g}
\def\kf{K_g}

\def\xiplus{\xi^+}
\def\ximinus{\xi^-}

\def\papertitlepage{\baselineskip 3.5ex \thispagestyle{empty}}
\def\Title#1{\vspace{1.5cm}\begin{center}
 {\Large\bf #1} \end{center} 
\vspace{1.5cm}}
\def\Authors#1{\begin{center} {\it #1} \end{center}}
\def\Abstract{\vspace{1.5cm}\begin{center} {\large\bf Abstract} 
           \end{center} \parbigskip}
\def\Komabanumber#1#2#3{\hfill \begin{minipage}{4cm} UT-Komaba #1
              \parn #2\parn #3\end{minipage}}

\renewenvironment{thebibliography}{\pagebreak[3]\par\vspace{0.6em}
\begin{flushleft}{\large \bf References}\end{flushleft}
\vspace{-1.0em}

\begin{enumerate}\if@twocolumn\baselineskip=0.6em\itemsep -0.2em
\else\itemsep -0.2em\fi\labelsep 0.1em}{\end{enumerate}}
\begin{document}
%%%%%%%%%%%%%%%%%%%%%%%%%%%%
\baselineskip=0.7cm
\papertitlepage
\vspace*{1cm}
\Komabanumber{93-13}{hepth@xxx/9310155}{October 1993}

%%%%%%%%%%%%%%%%%%%%%%%%%%%%
\Title{Extraction of Black Hole Geometry   \\
\vskip 1.5ex in Exactly Quantized \\ \vskip 1.5ex
Two Dimensional Dilaton Gravity} 
\Authors{{\sc Y.~Kazama
\footnote[2]{e-mail address:\quad  
kazama@tkyvax.phys.s.u-tokyo.ac.jp}
 \  and\ \ Y.~Satoh
% }  \\
\footnote[3]{e-mail address:\quad 
ysatoh@tkyvax.phys.s.u-tokyo.ac.jp} } \\
 \vskip 3ex
 Institute of Physics, University of Tokyo, \\
 Komaba, Tokyo 153 Japan \\
  }
\vspace{1.5cm}
%%%%%%%%%%%%%%%%%%%%%%%%%%%%%%%%%%%%%%%
\Abstract
\vspace{0.5cm}
%%%%%%%%%%%%%%%%%%%%%%%%%%%%%%%%%%%%%%
\baselineskip=0.7cm 
Based on our previous work, in which a model of two dimensional 
dilaton gravity of the type proposed by Callan, Giddings, Harvey and 
Strominger was rigorously quantized,  we explicitly demonstrate 
 how one can extract space-time geometry in exactly solvable 
 theory of quantum gravity.  In particular, we have been able 
 to produce a prototypical configuration in which a ( smeared ) 
matter shock wave generates a black hole without naked sigularity.
%%%%%%%%%%%%%%%%%%%%%%%%%%%%%%%%%%%%%
\\
%\\
%PACS number(s): 04.06.+n
\newpage
%%%%%%%%%%%%%%%%%%%%%%%%
\indent
Notably with the advent of a model proposed by Callan, Giddings, 
 Harvey and Strominger (CGHS)\cite{CG},  dilaton gravity in two dimensions 
has been widely recognized as an excellent arena in which to discuss 
a variety of fundamental issues in quantum gravity, especially the 
 ones concerning the  quantum properties of a black hole.  Indeed we 
 now have a large body of literature on the CGHS model and its 
 variants, either employing  semiclassical approximations \cite{SC} or 
 attempting at exact treatments \cite{ABC,HT,VV} .  Semiclasscal analysis 
 has an 
 advantage that it is readily interpretable, but it suffers from 
 the limitations that it is valid only for large black hole mass 
 and for a large number of matter fields.  In contrast, 
 models amenable to exact quantization are in principle
 free of such limitations  but in practice it is far more difficult 
 to interpret the results in physical terms. \parsmallskip
%%%%
The purpose of this letter is to fill this vexing gap between these two 
approaches.  Specifically, we demonstrate explicitly 
 how one can extract  space-time geometry in which a matter 
 shock wave generates a black hole in the framework of exact 
quantization developed in our previous work \cite{HKS} ( hereafter referred to
 as I).  This is accomplished by fixing a gauge within the 
 conformal gauge and computing the mean values of the matter 
 energy-momentum tensor and the (inverse) metric in a particular 
 type of coherent physical state.  
Although an attempt has recently been made \cite{AL}, we believe that 
this is  the first time that one can explicitly see the emergence of 
 space-time geometry in an exactly solvable model of quantum 
dilaton gravity containing matter fields.  In this article, 
 we shall describe the essence of our work.  The details
 will be elaborated elswhere \cite{KS2}. 
\parsmallskip
We begin with a brief review of the model and the main results 
 obtained in I. 
%%%%%%%%%
The classical action of our model is of the  CGHS form \cite{CG}, 
given by 
\eqabegin
    S &=& {1 \over \ff} \int d^2 \xi \sqrt{-g}\left\{\e^{-2{\phi}}\left[
    -4g^{\alpha\beta} \partial_\alpha\phi\partial_\beta\phi
       -R_g  +4\lambda^2  \right]\ 
      + \sum_{i=1}^N \half g^{\alpha\beta}\partial_
         \alpha f_i\partial_\beta f_i\right\} \comma
   \label{eqn:cghs}       
\eqaend
where $\phi$ is the dilaton field and  $f_i\ (i=1,\ldots, N) $ are  
N massless scalar fields representing matter degrees of freedom. 
 We take the signature of the metric to be $(+-)$
 \cite{note1}.
%\footnote{ Compared with the form we adopted in I, 
%the signs of the terms in the bracket $\left[
%\ \ \right]$ are reversed to conform to the original CGHS action. 
%This however results only in a reverse of sign for $\chi$ in Eq.(4).}
\parsmallskip
%%%%%%
To define all the quantities unambiguously, we take 
our universe to be spatially periodic with period $2\pi L$. It is then
convenient to introduce coordinates 
$x^\mu = (t, \sigma) = \xi^\mu / L $ and require that all the fields 
in the action be invariant under $\sigma \rightarrow \sigma + 2\pi $. 
When the action is rewritten in terms of $x^\mu$, it retains its form 
except with the replacement $\lambda \rightarrow \mu \equiv 
\lambda L $. For physical interpretation, we will get back to
 the original variables $\xi^\mu$ and $\lambda$. \parsmallskip
%%%%%%%%%%%%%%%%%
Quantization of this model enforcing  conformal invariance was 
proposed by several authors \cite{ABC,HT,VV} and we adapted 
the approach of Ref \cite{HT}.  In this scheme, 
 one makes a classical transformation  $\Phi \equiv \e^{-2\phi}$,
 and  then implements an appropriate functional measure following \cite{DDK}.  
The resulting  quantum model takes a particularly simple form for 
$N=24$   given by 
\eqabegin
 S &=& S^{cl} + S^{gh}\comma \\
 S^{cl} &=& {1\over \ff} \int d^2x \left( -\del_\alpha\Phi\del^\alpha\Phi
 - 2\del_\alpha\Phi\del^\alpha\rho +4\mu^2\e^{\Phi + 2\rho} 
 + \half \del_\alpha \vec{f} \cdot \del^\alpha \vec{f} \right) \comma
 \label{eqn:cl}
\eqaend
where $\rho$ is related to the metric and the dilaton field through 
$g_{\alpha\beta} = \e^{2\rho -2\omega}\eta_{\alpha\beta}$ 
with $2\omega = \ln\Phi -\Phi$ and   $S^{gh}$ is the usual 
$b$-$c$ ghost action. This is the model which 
we solved exactly in our previous work by means of a quantum 
canonical mapping into free fields. \parsmallskip
From the equations of motion, the dilaton field $\Phi$ and the 
\lq\lq Liouville " field $\rho$ can be expressed in terms of periodic 
free fields $\psi$ and $\chi$ as 
\eqabegin
 \Phi &=& -\chi -AB \comma \qquad 
 \rho = \half ( \psi -\Phi )\period \label{eqn:Phirho}
\eqaend
The functions $A(\xplus)$ and $B(\xminus)$ are defined by
\eqabegin
 \delplus A(\xplus)&=& \mu \e^{\psi^{+/2}(\xplus)} \comma 
\qquad 
 \delminus B(\xminus) = \mu\e^{\psi^{-/2}(\xminus)} \label{eqn:AB}
\comma
\eqaend
where $\psi^{\pm/2} (x^\pm)$ are the left- and right-
 going components of $\psi(x)$. ( The light-cone 
coordinates are defined as usual by $x^\pm = t\pm\sigma$. ) 
 With an abbreviation $\gt \equiv \gamma/\sqrt{4\pi}$, 
we write the Fourier mode expansion for $\psi$ as 
\eqabegin
\psi/\gt &=&  \qplus 
 + \pplus(\xplus +\xminus) + i\sum_{n\ne 0} 
 \left((\alpha^+_n / n)\e^{-in\xplus} 
+(\tilde{\alpha}^+_n / n) \e^{-in\xminus} \right) \comma
\eqaend
and similarly for $\chi/\gt$ and $f^i/\gt \equiv \phi^i_f $ 
with $(q^+, p^+, \alpha^+_n, 
 \tilde{\alpha}^+_n)$ replaced by $(q^-, p^-, \alpha^-_n, 
 \tilde{\alpha}^-_n)$ and $(q^i_f, p^i_f, \alpha^i_{f, n}, 
 \tilde{\alpha}^i_{f,n})$ respectively. 
The Solution for $A(\xplus)$ which satisfies the proper boundary 
condition  dictated by the behavior of the right hand side of the first 
equation  in (\ref{eqn:AB}) is given by 
( suppressing the $t$ dependence )
\eqabegin
 A(\sigma) &=& \mu C(\alpha) \int_0^{2\pi} d\sigma'
    \e^{(1/2)\epsilon(\sigma-\sigma')\ln \alpha}
 \e^{\psi^{+/2}(\sigma')} 
\comma\label{eqn:A}
\eqaend
while that for $B(\xminus)$ is similarly obtained by 
 changing the sign of the first exponential and replacing $\psi^{+/2}$
 by $\psi^{-/2}$ in (\ref{eqn:A}). 
In these expressions,  $\alpha = \exp(\gamma \sqrt{\pi}\pplus )$, 
$C(\alpha) =1/\left(\alpha^{1/2}-\alpha^{-1/2}\right)$ and  
$\epsilon(\sigma)$ is the usual stair-step function.
Note that we must require $\pplus$ not to vanish since otherwise 
 $C(\alpha)$ blows up. 
In terms of the free fields, the energy-momentum tensors $T_{\pm\pm}$ 
 take simple forms: 
\eqabegin
  \gt^2 T_{\pm\pm} &=&  
\del_\pm \chi \del_\pm \psi -\del_\pm ^2\chi
   + \half (\del_\pm \vec{f})^2  \period \label{eqn:fEMT}
\eqaend
\indent
It was shown in I  that the mapping from the 
original fields $\{\Phi, \rho\}$ into the free fields 
$\{\psi, \chi\}$ is a quantum as well as classical canonical 
transformation  if we assume the non-vanishing commutators among the 
modes of $\psi$ and  $\chi$ to be 
 $\left[q^\pm, p^\mp\right] = i\hbar$, 
 $\left[\alpha^\pm_m, \alpha^\mp_n\right] = 
 \left[\tilde{\alpha}^\pm_m, \tilde{\alpha}^\mp_n\right] = m\hbar 
 \delta_{m+n, 0}$.
%%%%%%%%%%
Commutation relations for the matter fields are of the usual form. 
\parsmallskip
 %%%%%
 The  model so quantized is conformally invariant with the central 
charges $c^{dL}=2$ and $c^f=24$ for the dilaton-Liouville (dL) and 
 the matter (f) sectors.  $\chi$ and  the product 
$A(\xplus)B(\xminus)$ tranform as 
genuine dimension  zero primary fields,  while due to the presence of
 the background charge   $\psi$ transforms anomalously as 
 $\left[L^{dL}_m, \psi(x)\right] = \hbar \e^{im\xplus}((1/i) \delplus
 \psi + m ) $.  \parsmallskip
%%%%%%%%%%
The mathematical structure of our model is very similar to 
that of a  bosonic string theory  
and all the physical states can be readily obtained through 
BRST analysis \cite{BMP,HKS}. They are of the generic form 
$ \ket{\Psi} = \ket{\Psi_0} + \ket{\Lambda}$, 
where $\ket{\Psi_0}$ is a special representative of a non-trivial 
 cohomology class satisfying $L^{tot}_0 \ket{\Psi_0}=
\bar{L}^{tot}_0 \ket{\Psi_0} =0 $ and $\ket{\Lambda}$ is the BRST 
 trivial part. 
%%%
The interpretation of physical states, however, is quite
 different from that in a string theory: In the present model, 
the  Virasoro level specifies the discretized energy carried by  
 a state. As we will be most interested in  configurations where the 
matter fields carry finite energy in the limit of large $L$, 
 physical states at arbitrary high Virasoro levels will 
 be of utmost importance. \parsmallskip
%%%%%%
To deal with such states,  so called DDF representation \cite{DDF} is
most useful.  For our model, a convenient set of 
 spectrum-generating DDF oscillators can be constructed as
 \cite{note2}
%\footnote{
%These oscillators are intimately related to those introduced in 
%\cite{VV}.}
%
\eqabegin
 \At^i_{-n} &\equiv & \e^{-i(n /\gt \pplus)
\ln(\gt \pplus)}
\int_0^{2\pi}{d\yplus \over 2\pi} \e^{-in
 \etap/(\gt\pplus)} \delplus\phi^i_f(\yplus) \comma \label{eqn:Atil}
\eqaend
where
\eqabegin
\etap &\equiv& \ln ( \exp(\gt \qplus /2) A(\xplus)/\mu) \period
\eqaend
$\etap$ consists only of the modes of $\psi^{+/2}
(\xplus)$ and it can be checked to be a genuine primary field of 
dimension $0$.  Physical states can be built by applying $\At^i_{-n}$'s
 on a zero-mode vacuum $\ket{\vec{P}}$, $( \vec{P} \equiv ( \pplus, 
\pminus, \vec{p}_f))$,  satisfying 
 the zero-energy condition  
$\pplus\pminus +(1/2)\vec{p}_f^2 -\hbar=0$.
It can be shown\cite{KS2} that the physical states so constructed agree 
precisely with the ones previously obtained in BRST formalism. 
\parsmallskip
%%%%%%%%%%%%%%%%%%%%%%%%%%%%
We shall now construct a  physical state, which will be seen to 
 describe a universe in which a smeared 
 matter shock wave produces a black hole.  
Physical meaning of such an 
abstract state can only be extracted by looking at its response to the 
action of appropriate operators of physical significance and  in this 
 article we shall use the left-going energy-momentum tensor of 
 the matter field $T^f(\xiplus)$ and the inverse metric 
$g^{\alpha\beta}$. They are 
expressed in terms of the free fields as 
\eqabegin
 T^f(\xiplus) &=& {1\over \gt^2 }:\left(\del_{\xiplus}
\vec{f}\right)^2 :\comma \qquad 
g^{\alpha\beta} = -\left(\nchiepsi + AB\e^{-\psi}\right)
\eta^{\alpha\beta} \period
\eqaend
The normal-ordering for $T^f(\xiplus)$ is the usual one while
the one for $\nchiepsi$ is defined by 
$ \nchiepsi = \chi \e^{-\psi} -\left[\chia, \e^{-\psi}\right]$, 
where $\chia$ stands for the annihiliation part of $\chi$. 
 As all the modes of $\psi$ 
 commute with each other, $AB\e^{-\psi}$ is well-defined without 
 normal-ordering.  $g^{\alpha
\beta}$  so defined is finite and hermitian. \parsmallskip
%%%%
Clearly the operators we have chosen to work with are  gauge 
dependent and hence we must specify $\ket{\Lambda}$ as well as 
 $\ket{\Psi_0}$. There are a number of general guide lines for 
choosing them.  As for $\ket{\Psi_0}$, 
we expect that 
it should be a  suitable coherent state 
 since we wish to obtain such macroscopic configuration as that of 
 a shock wave producing a black hole. On the other hand, a simple 
 argument on energy-momentum conservation shows that in order for the 
 mean values to exibit non-trivial coordinate dependence we must take 
 $\ket{\Lambda}$ to be a superposition of states  
with various Virasoro weights. \parsmallskip
%%%%%%%%%%%%%%%%%%%%%%%%%%%%%%%%%%%%%%%
Accordingly, we have chosen our physical state to be of the following
 form. First, the gauge part $\ket{\Lambda}$ is chosen 
to be of a simple form
\eqabegin
\ket{\Lambda} &=& {1\over \hbar^2} \left(d\, b_{-1}
 + \dbar\, \bbar_{-1}\right)\Omegaket \comma
\eqaend
where $b_{-1}$ and $\bbar_{-1}$ are, respectively, the left- and 
right-going anti-ghost oscillator at level one. 
 $\Omegaket$ is a superposition of zero-mode states of 
the form
\eqabegin
\Omegaket & \equiv & \sum_{k= - \infty }^{\infty} \omega_k 
\sum_{l= \pm 1 ,0}
           \mid \tilde{P}(k,l)> \comma  \label{eqn:Omega} \\
\mid \tilde{P}(k,l)>  & \equiv & \, e^{-ic p^+/\hbar \gamma}
  \gamma^2 \int_{-\infty}^{\infty} dp^+ 
  \int_{- \infty}^{\infty} dp^- \, \weight \mid p^+,p^-(k,l),\vec{p}_f 
  > \nn\\
 & & \times  \delta (p^- - \frac1{p^+}(\hbar - \half p_f^2)) 
\comma  \label{eqn:Pkl} \\
      p^- (k,l) & \equiv & p^- - \frac{\hbar}{p^+} k  - 
       i \hbar \gt\, l \comma   
\eqaend
%%%%%%%%%%%%%%%%%%%
where $\omega_k$ are a set of real coefficients. 
 In $\mid \tilde{P}(k,l)>$, smearing with an 
 appropriate real weight $\weight$, to be specified later, is introduced 
 to make the mean value of  the operator $\qminus$ well-defined, 
which  appears in $<\, g^{\alpha\beta}\, >$. Further 
a BRST invariant phase factor in front produces a 
coordinate-independent contribution in $< g^{\alpha\beta}> $ and 
the constant $c$ will  be adjusted to cancel certain unwanted terms. 
Summation over states with shifted $\pminus$ zero-modes in 
(\ref{eqn:Omega}) is necessary for obtaining non-trivial 
expectation values. 
\parsmallskip
%%%%%%%%%%%%%%%%%%%%%%%%%%%%
As for the BRST non-trivial part $\ket{\Psi_0}$, we have taken it to be 
\eqabegin
\Psizeroket & \equiv & e^G \psmearvac \comma  \qquad 
G \equiv  {1\over \hbar}\sum_{n\ge 1}{ \tilde{\nu}_n\over n}
\Atil_{-n} \comma \\
\tilde{\nu}_n &=& \nu_n \e^{in x^+_0 }
\qquad (\nu_n,\, x^+_0\,:\, \mbox{real constants}) \comma
\eqaend
where $\psmearvac \equiv \mid \tilde{P}(0,0)>$ defined in 
 (\ref{eqn:Pkl}). It is easy to see that $L^{tot}_0\psmearvac = 
 \bar{L}^{tot}_0\psmearvac =0$ holds. 
 For simplicity, we consider a coherent state in which only one kind
 of  matter field is excited, and thus omit the superscript $i$ on 
 $\tilde{A}_{-n}$. The phase factor in 
$\tilde{\nu}_n$ will produce a  shift 
$\xplus \rightarrow \xplus -x^+_0$ in certain terms  
  and  will specify where a matter shock wave will 
traverse. \parsmallskip
%
%%%%%%%%%%%%%%%%%%%%%%%%%%%%%%%%%%%%%%%%%%%%%
We are now ready to describe the computations of the mean values.
After a simple calculation, the mean value of a hermitian operator 
$\calO$ without ghosts can be reduced to the form 
\eqabegin
 \bra{\Psi}\calO\ket{\Psi} &=& 
 \bra{\Psi_0}\calO\ket{\Psi_0} \nn\\
&& +
(1/ \hbar) \Omegabra
 \left[ \calLp, \calO \right] \ket{\Psi_0}
+(1/ \hbar) \bra{\Psi_0} \left[\calO,
 \calLm \right] \Omegaket  \\
&& +  (1 / \hbar^2)\Omegabra
  \calO \left[\calLp, \calLm\right] \Omegaket 
+ (1 / \hbar^2)\Omegabra \left[ \left[ \calLp, \calO\right], 
\calLm\right] \Omegaket \comma \nn
\eqaend
where 
$\calLm \equiv \Ltot_{-1} + \Lbartot_{-1}$ and 
$\calLp \equiv \Ltot_{1} + \Lbartot_{1}$.
%%%%%%%%%%%%%%%%%%%%%%%%
The rest of the calculations, although tedious, can be performed 
 without approximation in a straightforward way
\cite{note3}.
%\footnote{ There is a techinicality in the definition of the inner product 
 %for the zero-mode sector, but this can be settled by demanding that 
 %$<\,T^f\,>$ and $<\,g^{\alpha\beta}\,>$ be real. For 
 %details, see \cite{KS2}.}. 
The exact results are unfortunately 
long and involved,  and we shall display them elswhere \cite{KS2}. 
Here instead, we present the most interesting 
limit, namely when the 
(parameter) size of the universe, $L$, becomes very large.  In this 
 limit, as the discrete energy levels tend to become continuous, 
 we can replace certain infinite sums by integrals and the results 
simplify  considerably. \parsmallskip
%%%%%%%%%%%%%%%%%%%%%%%
It turns out that the following choice of parameters will produce 
 a shock-wave-black-hole configuration:
\eqabegin
 \nu_n &=& \nu(Lu) = \nu u^{-1/2} \e^{-au^2} \comma 
 \qquad p_f=0\comma\\ 
 \omega_n &=& \omega(Lu) = -\omega/ (Lu) \qquad (n \ne 0, \ \omega < 0)
 \comma\\
 \weight &=& \pplus \e^{-\alpha(\pplus-\pplus_0)^2/2} \comma
\eqaend
where we have introduced a quantity $u\equiv n/L$, 
 to be used as an integration variable, and 
$\nu$, $\pplus_0$, $\omega$ and $\omega_0$  are constants. 
For these parameters, large $L$ limit of the mean values become 
\eqabegin
<\,T^f (\xi^+) \,>&\stackrel{L \rightarrow \infty}{\sim}&
    \omega \nu^2 I_T(\xiplus -\xiplus_0 )
  \, \psmearnorm    \comma \\
%%%%%%%%%%%%%
<\, g^{-1} \, > &\equiv & - <\, : \left( \chi+AB\right)\e^{-\psi} :\,> 
 \nn\\ 
 &\stackrel{L \rightarrow \infty}{\sim}&
 \left( -\kf(\xiplus) -\lm^2 \xiplus\ximinus \right) \psmearnorm \\
 \kf(\xiplus) &=& c_K\xiplus + d_K -\gt^2\omega\nu^2 I_\chi(\xiplus
 -\xiplus_0) \comma \label{eqn:ginv} 
\eqaend
where  
 $I_T(\xi)$ and $I_\chi(\xi)$ are integrals of the form 
\eqabegin
 I_\chi(\xi) &=& \int_{1/L}^\infty du {\cos u\xi \over u^2}
\int_0^u dv \left[v(u-v)\right]^{-1/2} \e^{-a(v^2 + (u-v)^2)} \comma
 \label{eqn:IKdef}  \\
 I_T(\xi) &=& -\del_\xi^2 I_\chi(\xi) \label{eqn:IT} \period
\eqaend
$\lm^2$ is a certain constant, proportional to $\lambda^2$, which 
 is a function of the parameters $M$, $\omega$, $\omega_0$, $\alpha$
 and $p^+_0$, and the norm $\psmearnorm$ is a function of $\alpha$
 and $p^+_0$ times a $\delta$-function $\delta^{25}(0)$. We shall 
 ignore this common factor, which can be made finite by additional 
smearing.  The term of the form $-\lm^2\xiplus\ximinus$ 
arises  entirely from the pure gauge part of $<\, AB\e^{-\psi}\,>$, 
 after a suitable adjustment of $\omega_0$, and it describes  the 
 so called linear dilaton vacuum when the matter field vanishes. The 
 fact that our simple choice of $\ket{\Lambda}$ produces 
 precisely such a vacuum configuration is quite remarkable. 
In $\kf$, the constants $c_K$ and $d_K$ can be made arbitrary by 
adjusting $\pplus_0$, $\alpha$ in $\weight$ and  the 
 constant $c$  introduced in the definitions of $\ket{\tilde{P}(k,l)}$
 and $ \psmearvac $. \parsmallskip
%%%%%
By a change of variable, the integrals $I_T$ and $I_\chi$ can be 
 brought to the following forms: 
\eqabegin
 I_T &=& \sqrt{{2\pi\over a}}\int_0^1 dx (1-x^2)^{-1/2}(1+x^2)^{-1/2}
 \e^{-\xi^2/(2a(1+x^2))} \comma \label{eqn:ITmhalf} \\
 I_\chi &=& L\pi  -\sqrt{2\pi a}\int_0^1 dx
\cx{-1/2}{1/2} \e^{-\xi^2/(2a(1+x^2))} \nn\\
 && \quad -\sqrt{{2\pi \over a}}\xi^2 \int_0^1 dx
 \cx{-1/2}{-1/2} \Fii\left( \half; {3\over 2};
 -{\xi^2 \over 2a(1+x^2)} \right)  \\
\xilarge L\pi -{\pi^2\over 2}|\xi| \comma \label{eqn:IKmhalf} 
\eqaend
where $\Fii(a;b;z)$ is the confluent hypergeometric function. 
It is evident from (\ref{eqn:ITmhalf}) that 
 $<\, T^f(\xiplus)\,>$ is very nearly  a Gaussian peaked around 
$\xiplus_0$ and as the parameter $a$  approaches $0$ it becomes a 
$\delta$-function. Thus it precisely describes a (smeared) 
 shock wave of left-going matter 
energy density  discussed in \cite{CG}. 
 As for $I_\chi(\xiplus-\xiplus_0)$, it is important to note that 
 the asymptotic behavior
 for large $|\xiplus -\xiplus_0|$ is {\it linear} as is seen in 
(\ref{eqn:IKmhalf}). In fact a numerical evaluation shows 
 that for small $a$  this asymptotic form is accurate even for 
 small $|\xiplus -\xiplus_0|$ down to $\sim \sqrt{a}$.  Thus we can make 
 $<\, g^{-1}\,>$ behave very much like the CGHS case by adjusting 
 the term $c_K\xiplus +d_K$ in $\kf$ to cancel this linear portion for 
the range $\xiplus < \xiplus_0$ and to remove the constant $L\pi$. 
  $\kf(\xiplus)$ then becomes 
\eqabegin
 \kf(\xiplus) &=& \gt^2\omega \nu^2 \left( {\pi^2\over 2}
 (\xiplus -\xiplus_0) -I_\chi(\xiplus -\xiplus_0)+L\pi \right) \comma
\label{eqn:kf}
\eqaend
which behaves like $ (\pi^2/2)\gt^2\omega \nu^2
\theta(\xiplus -\xiplus_0) $ for $|\xiplus -\xiplus_0|>>\sqrt{a}$. 
In this way we obtain a smeared version of the CGHS black hole.
In Fig. 1, we show the line of curvature singularity for our 
configuration obtained numerically. 
 We clearly see that a black hole without a naked 
 singularity is formed and to the left of 
the line $\xiplus=\xiplus_0$ the space-time quickly becomes the 
 linear dilaton vacuum configuration for $\sqrt{a} << \xiplus_0$.
 Accordingly, for small enough 
 $a$, the black hole mass takes exactly the CGHS form, which in 
 our normalization is given by $(\pi^2/2)\lm \omega\nu^2\xiplus_0$. 
\parsmallskip
%%%%%
By using an exactly solvable model of CGHS type, we have shown 
explicitly how one can reproduce a black hole  geometry from an 
exact yet abstract physical state in quantum theory of gravity. 
Although the model employed is but a toy model in $1+1$ dimensions, 
we believe that it is quite significant to be able to discuss one of the 
  important issues in quantum gravity in a concrete and 
unambiguous manner. Furthermore, the point of view and the procedures 
developed in this work should find wide applications in other models of 
quantum gravity and possibly in quantum cosmology. \parsmallskip
%%%%%%%%%%%%%%%%%%%%%%%%
We would like to thank our colleagues, especially S. Hirano and 
 H. Ishikawa, for discussions. 
 The research of Y.K. is supported in part 
by the Grant-in-Aid for Scientific Research (No.04640283) 
 and Grant-in-Aid for scientific Research for Priority Areas  
(No. 05230011) from the Ministry of Education,Science and Culture.
%%%%%%%%%%%%%%%%%%%%%%%%%%%%%%%%%%%
%%%%  references  %%%%%%

%%%%%%%%%%%%%%%%%%%%%%%%%%%%
%%% figure %%%
\newpage
\epsfbox{figure1.ai}
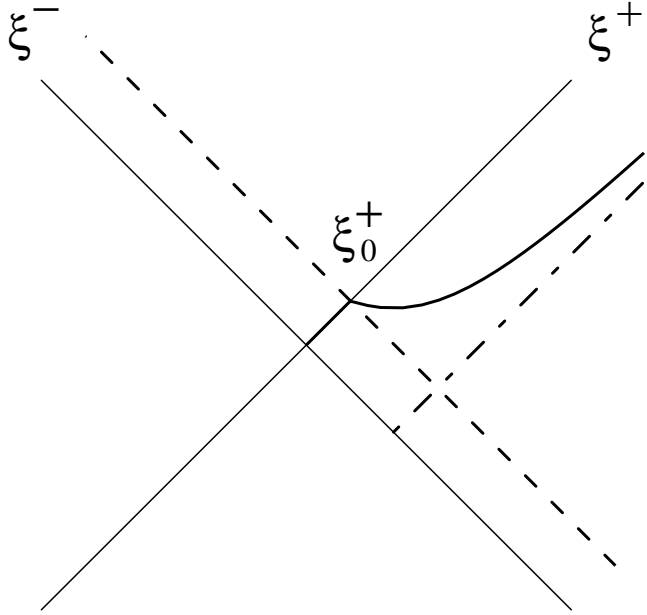
{\bf Fig.1 } : The line of curvature singurarity ( solid line )  
        produced by a left-going smeared shock wave along 
        $ \xi^+ = \xi^+_0 $ ( dotted line ). The dot-dashed line represents 
        the event horizon and the space-time quickly approaches the linear 
        dilaton vacuum to the left of $ \xi^+ = \xi^+_0 $. \\
%%%%%%%%%%%%%%%%%%%%%%%%%%%%

\begin{thebibliography}
\bibitem{note1} Compared with the form we adopted in I, the sign of the terms 
                in the bracket [ ] are reversed to conform to original CGHS 
                action. This however results only in a reverse of sign for 
                $ \chi $ in Eq.(4).
\bibitem{note2} The oscillators are intimately related to those introduced 
                in \cite{VV}.
\bibitem{note3} There is a technicality in the definition of the inner product 
                for the zero-mode sector, but this can be settled by demanding 
                that $ < T^f > $ and $ g^{\alpha \beta} $ be real. For details,
                see \cite{KS2}.
\end{thebibliography}
\end{document}